\documentclass[aps,prb,showpacs,twocolumn,amsmath,amssymb,superscriptaddress]{revtex4}
\usepackage{graphicx}
\usepackage{Jonasmacros}
\usepackage{bm}
\begin{document}
\date{\today}
\title{Dynamical properties of a vibrating molecular quantum dot in a Josephson junction}
\author{J. Fransson}
\email{Jonas.Fransson@fysik.uu.se}
\affiliation{Department for Physics and Astronomy, Box 530, SE-751 21\ \ UPPSALA, Sweden}
\author{A. V. Balatsky}
\affiliation{Theoretical Division, Los Alamos National Laboratory, Los Alamos, New Mexico 87545, USA}
\affiliation{Center for Integrated Nanotechnology, Los Alamos National Laboratory, Los Alamos, New Mexico 87545, USA}
\author{Jian-Xin Zhu}
\affiliation{Theoretical Division, Los Alamos National Laboratory, Los Alamos, New Mexico 87545, USA}

\begin{abstract}
We investigate dynamical transport aspects of a combined nanomechanical-superconducting device in which Cooper pair tunneling interfere with the mechanical motion of a vibrating molecular quantum dot embedded in a Josephson junction. Six different regimes for the tunneling dynamics are identified with respect to the electron level and the charging energy in the quantum dot. In five of those regimes new time-scales are introduced which are associated with the energies of the single electron transitions within the quantum dot, while there is one regime where the internal properties of the quantum dot are static.
\end{abstract}
\maketitle

\section{Introduction}
\label{sec-intro}
Inelastic scattering processes carry dynamical degrees of freedom which have a large influence on the electron dynamics. Signatures in the conductance of molecular electronics devices \cite{langlais1999,park2000,smit2002,park2002,reichert2002,zhitenev2002} indicate coupling between electronic and vibrational degrees of freedom, whereas spin inelastic scattering effects have been utilized in experimental studies of the magnetic properties of magnetic atoms and clusters.\cite{gambardella2003,heinrich2004,hirjibehedin2006,hirjibehedin2007,fransson2009,balashov2009} A huge effort has been spent on studies of the influence of spin and vibrational degrees of freedom on the transport through quantum dots (QDs).\cite{rozhkov1999,flensberg2003,zhu2004,siano2004,franssonPRB2008,galperin2008} Studies of various aspects of inelastic scattering effects are of fundamental importance.

Recent developments towards incorporating superconducting electronics into nanoelectromechanical devices open possibilities of cooling \cite{naik2006,pekola2007} and Cooper pair shuttling.\cite{gorelik2001,zazunov2006,zhu2006,fransson2008} In this paper we focus on the dynamical aspects of a QD embedded in a Josephson junction, which to our knowledge not has been studied previously.

\begin{figure}[b]
\begin{center}
\includegraphics[width=0.45\textwidth]{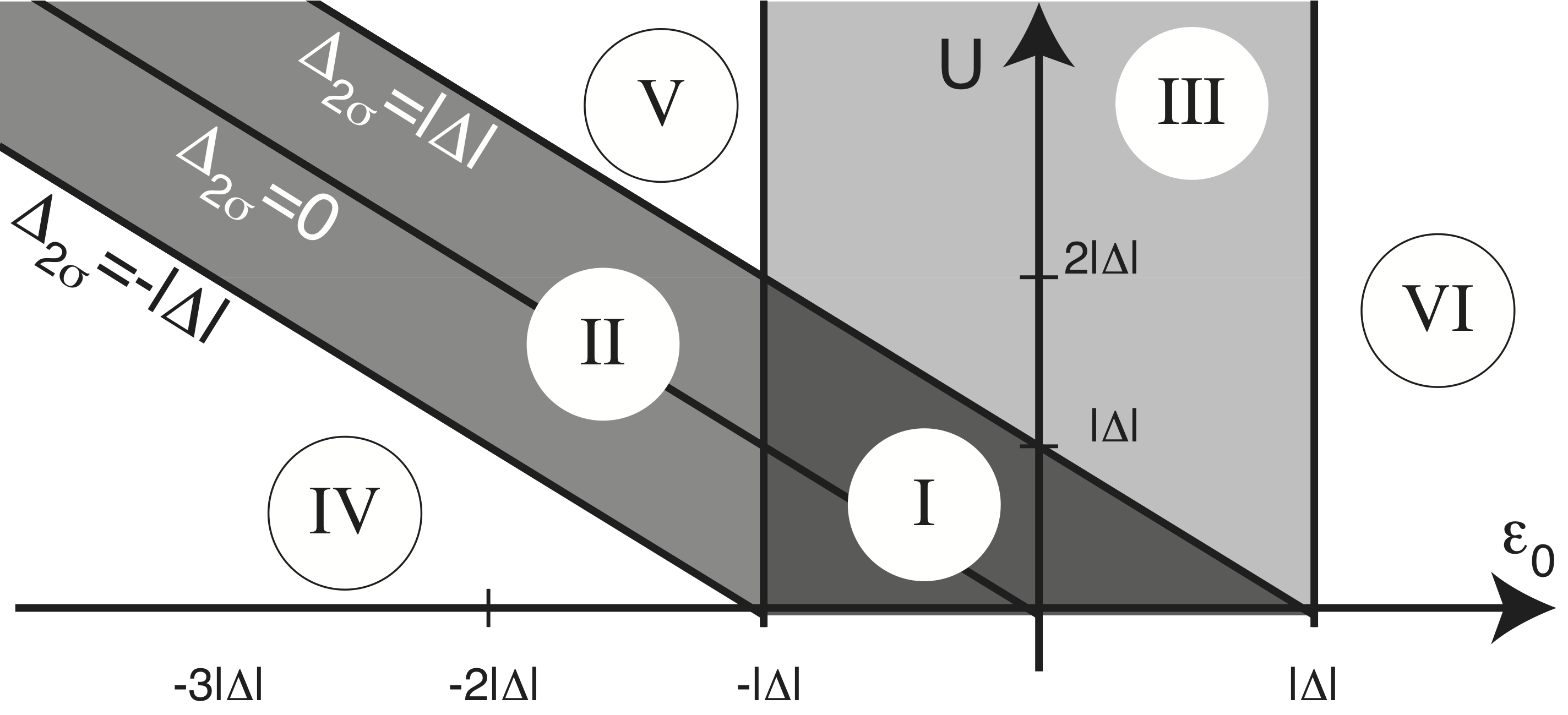}
\end{center}
\caption{Phase diagram of the dynamics of the molecular QD, embedded in a Josephson junction, with respect to the energy level $\dote{0}$ and charging energy $U$. Here, $\Delta_{pq}=E_p-E_q$ and $(E_0,E_\up,E_\down,E_2)=(0,\dote{\up},\dote{\down},\dote{\up}+\dote{\down}+U)$, where $\dote{\sigma}=\dote{0}+\sigma B/2$, with $B=0$, whereas $|\Delta|$ is the superconducting gap in the electrodes.}
\label{fig-pd}
\end{figure}

In this paper, we consider the influence of the mechanical motion of a molecular QD, embedded in a Josephson junction, on the supercurrent flowing across the junction. The mechanical motion of the QD couples to the electronic degrees of freedom of the tunneling electrons, which dramatically influence the dynamics of the electronic occupation in the QD. We can clearly distinguish between six different regimes, with respect to the electron level $\dote{0}$ and charging energy $U$ in the QD, see Fig. \ref{fig-pd}, in which the dynamical aspects of the transport properties are different. In regimes II | VI, the dynamics of the QD properties generate new time-scales to the transport which are intimately associated with the energies of the single electron transitions in the QD. In regime I, on the other hand, the internal properties of the QD are static, even for finite bias voltages. The motion and Josephson current are, thus, set solely by the applied bias voltage and phase difference between the electrodes.

The paper is organized as follows. In Sec. \ref{sec-model} we introduce the model of the Josephson junction in which the molecular QD is embedded, and we derive the basic expressions for the Josephson current, the two-electron tunneling process, and the occupation numbers in the QD. The results are analyzed in Sec. \ref{sec-dynamics} and we conclude the paper in Sec. \ref{sec-discussion}.

\section{Model of the Josephson current}
\label{sec-model}
\begin{figure}[t]
\begin{center}
\includegraphics[width=0.45\textwidth]{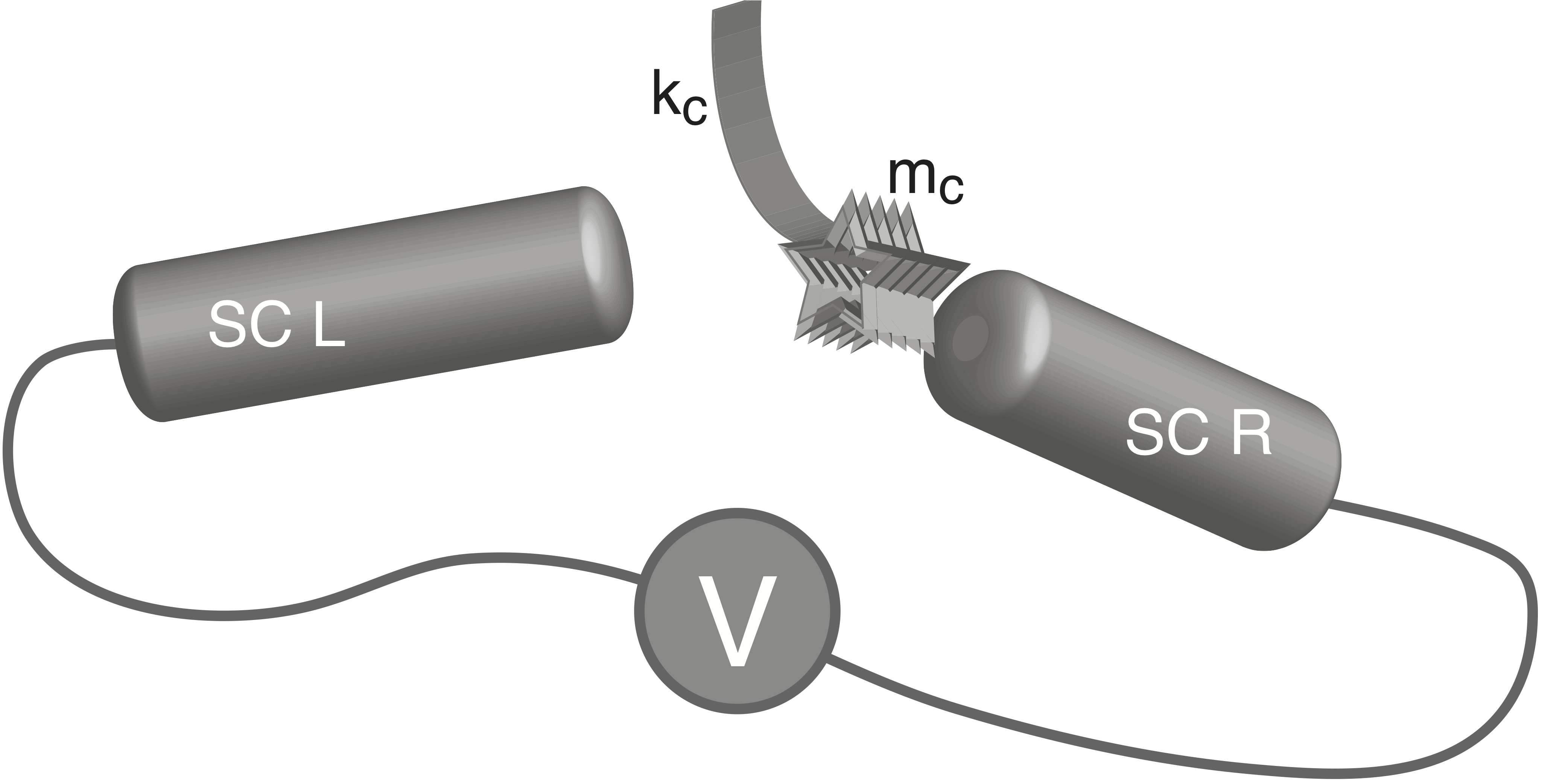}
\end{center}
\caption{Schematic view of the mechanically and electronically coupled QD to the superconducting electrodes (SC L) and (SC R). The QD is suspended on a cantilever which is modeled as a harmonic oscillator with spring constant $k_c$ and mass $m_c$. The device is biased with the voltage $V$.}
\label{fig-system}
\end{figure}

We start by considering a molecular QD embedded in a Josephson junction, where the QD is exposed to mechanical oscillations which are modeled with a Hook's law constant $k_c$, and we assume that there is a bias voltage applied across the junction. We illustrate the system schematically in Fig. \ref{fig-system}.

The Hamiltonian for the set-up is expressed by
\begin{align}
\Hamil=\Hamil_L+\Hamil_R+\Hamil_{QD}+\Hamil_T,
\end{align}
where $\Hamil_\chi=\sum_{\bfk\sigma\in\chi}\dote{\bfk}\cdagger{\bfk}\cc{\bfk}+\sum_{\bfk\in\chi}(\Delta\csdagger{\bfk\sigma}\csdagger{-\bfk\bar\sigma}+H.c.)$, $\chi=L,R$, are usual $s$-wave BCS Hamiltonians, whereas $\Hamil_{QD}=\sum_\sigma\dote{\sigma}\ddagger{\sigma}\dc{\sigma}+Un_\up n_\down$ defines the QD, with single-electron levels $\dote{\sigma}=\dote{0}+\sigma B/2$ which are spin split by the effective field $B$, and with charging energy $U$. Finally,  $\Hamil_T=\sum_{\bfk\sigma}t_\chi(\cdagger{\bfk}\dc{\sigma}+H.c.)$ accounts for the single-electron tunneling between the lead $\chi$ and the QD with rate $t_\chi$. The local vibrational mode of the island is in the linear coupling regime given by
\begin{align}
t_L=t_L^{(0)}(1+\alpha_L q),\quad t_R=t_R^{(0)}(1+\alpha_R q),
\end{align}
where $\alpha_{L(R)}$ describes the coupling between the tunneling electrons and the vibrational mode corresponding the to left (right) tunnel junction. The quantity $q$ is the displacement operator for the oscillator. The tunneling matrix element $t_\chi$ is exponential in the displacement $q$, thus, the assumed linear coupling is a good approximation for small $q$. This allows evaluation of $\alpha_{L(R)}$ in terms of the tunneling matrix elements and their distance dependence. We assume here a very general equilibrium geometry with no particular symmetry being required. The equilibrium position ($q=0$) for the QD within the junction may be placed anywhere in between the leads.

\subsection{Derivation of the current}
\label{ssec-current}
We derive the basic formula for the supercurrent through the system using an analogous procedure as in Refs. \onlinecite{zhu2006,fransson2008}. The total current flowing between the QD and lead $\chi$ can be fundamentally expressed as $I_\chi^{(\tot)}(t)=-e\dt\av{\sum_{\bfp\sigma\in\chi}\cdagger{\bfk}\cc{\bfk}}$, from which we obtain
\begin{align}
I_\chi^{(tot)}=&
	2e\re\int_{-\infty}^t
		\biggl(
			\av{\com{A_\chi(t)}{A^\dagger_\chi(t')}}e^{i\mu_\chi(t-t')}
\nonumber\\&
			+\av{\com{A_\chi(t)}{A_\chi(t')}}e^{i\mu_\chi(t+t')}
		\biggr)
	dt'.
\end{align}
Here, we have defined the current operator 
\begin{align}
A_\chi(t)=\sum_{\bfk\in\chi,\sigma}t_\chi(t)\cdagger{\bfk}(t)\dc{\sigma}(t),
\end{align}
using that $\cdagger{\bfk}(t)=e^{i(\Hamil_\chi-\mu_\chi N_\chi)t}\cdagger{\bfk}e^{-i(\Hamil_\chi-\mu_\chi N_\chi)t}$ and analogously for the $d$-operators, where $\mu_\chi$ and $N_\chi=\sum_{\bfk\sigma\in\chi}\cdagger{\bfk}\cc{\bfk}$ is the chemical potential and total number of electrons in lead $\chi$, whereas we put the chemical potential for the QD to zero, i.e. $\mu_{QD}=0$.

The total current comprise one component of single-particle, or electron, current which is contained in the first term of the above expression, while the second term describes the supercurrent. In the present paper, we are interested in the supercurrent only and, therefore, we discard the first contribution to the total current in the following discussion.

The average $\av{\com{A_\chi(t)}{A_\chi(t')}}$ is decoupled into the averages e.g. $\av{\cdagger{\bfk}(t)\csdagger{\bfk\bar\sigma}(t')}$ and $\av{\dc{\sigma}(t)\dc{\bar\sigma}(t')}$ of which the former is further handled via e.g. Bogoliubov-Valatin transformation $\cc{\bfk}=u_\bfk\gamma_{\bfk\sigma}-\sigma v_\bfk^*\gamma_{\bfk\bar\sigma}^\dagger$, such that ($\tau=t-t'$)
\begin{subequations}
\begin{align}
\calF^{\dagger,>}_{\bfk\sigma\bar\sigma}(t,t')\equiv&
	(-i)\av{\cdagger{\bfk}(t)\csdagger{\bfk\bar\sigma}(t')}
\nonumber\\=&
	i\sigma u_\bfk^*v_\bfk
		[
			f(-E_\bfk)e^{-iE_\bfk\tau}-f(E_\bfk)e^{iE_\bfk\tau}
		],
\\
\calF^{\dagger,<}_{\bfk\sigma\bar\sigma}(t,t')\equiv&
	i\av{\csdagger{\bfk\bar\sigma}(t')\cdagger{\bfk}(t)}
\nonumber\\=&
	i\sigma u_\bfk^*v_\bfk
		[
			f(-E_\bfk)e^{iE_\bfk\tau}-f(E_\bfk)e^{-iE_\bfk\tau}
		].
\end{align}
\end{subequations}
Here, $E_\bfk=\sqrt{(\dote{\bfk}-\mu_\chi)^2+|\Delta_\chi|^2}$ defines the quasi-particle energies, whereas $u_\bfk=\sqrt{(1+[\dote{\bfk}-\mu_\chi]/E_\bfk)/2}$ and $v_\bfk=\sqrt{(1-[\dote{\bfk}-\mu_\chi]/E_\bfk)/2}$, which satisfy $|u_\bfk|^2+|v_\bfk|^2=1$ and $u_\bfk^*v_\bfk=|\Delta_\chi|e^{i\phi_\chi}/(2E_\bfk)$, with the macroscopic phase $\phi_\chi$. We shall proceed at low temperatures, such that we can approximate $f(E_\bfk)\approx0$ and $f(-E_\bfk)\approx1$.

The average $\av{\dc{\sigma}(t)\dc{\bar\sigma}(t')}$ is handled by transforming the QD Hamiltonian into diagonal form using the expansion $\dc{\sigma}=\X{0\sigma}{}+\sigma\X{\bar\sigma2}{}$ for the operators $\X{pq}{}=\ket{p}\bra{q}$, $\Delta_{pq}=E_p-E_q$, $p,q\in\{0,\sigma,2\}$.\cite{fransson2005}, such that $\Hamil_{QD}=\sum_{p=0,\sigma,2}E_p\X{pp}{}$, $(E_0,E_\up,E_\down,E_2)=(0,\dote{\up},\dote{\down},\dote{\up}+\dote{\down}+U)$. Then, since $\dc{\sigma}(t)\dc{\bar\sigma}(t')=(\X{0\sigma}{}+\sigma\X{\bar\sigma2}{})(t)(\X{0\bar\sigma}{}+\bar\sigma\X{\sigma2}{})(t')$, we define the anomalous averages for the QD according to
\begin{subequations}
\begin{align}
F_{\bar\sigma\sigma}^>(t',t)\equiv&
	(-i)\av{\dc{\bar\sigma}(t')\dc{\sigma}(t)}
%\nonumber\\=&
	=(-i)\bar\sigma N_{02}e^{i\Delta_{\bar\sigma0}\tau},
\\
F_{\bar\sigma\sigma}^<(t',t)\equiv&
	i\av{\dc{\bar\sigma}(t')\dc{\sigma}(t)}
%\nonumber\\=&
	=i\bar\sigma N_{02}e^{i\Delta_{2\sigma}\tau},
\end{align}
\end{subequations}
where $N_{02}=\av{\X{0\sigma}{}\X{\sigma2}{}}=\av{\X{02}{}}$ is the average rate for the two-electron transition $\X{02}{}\equiv\ket{0}\bra{2}$ at the energy $\Delta_{20}=E_2-E_0$.

Using the prescribed procedure, we find that the supercurrent between the superconducting lead $\chi$ and the QD at low temperatures can be written 
\begin{align}
I_\chi(t)=&
	-\re\int N_{02}[J_\chi(\mu_L)(1+\alpha_\chi q)^2\sin(\omega_\chi t+\phi_\chi)
\nonumber\\&
				-\Gamma_\chi(\mu_\chi)(1+\alpha_\chi q)\alpha_\chi\dot{q}\cos(\omega_\chi t+\phi_\chi)]
				e^{-i\omega\tau}dt'\frac{d\omega}{2\pi},
\label{eq-ISmain}
\end{align}
where $\omega_\chi=2\mu_\chi$, and where we have assumed the local approximation $t_\chi(t')\simeq t_\chi(t)-\tau\dot{t}_\chi(t)$, which is justified since the vibrational motion is much slower than the electronic tunneling processes. The amplitudes $J_\chi$ and $\Gamma_\chi$ of the Josephson current in absence and presence of the coupling to the vibrational mode are given by
\begin{subequations}
\begin{align}
J_\chi(\mu_\chi)=&2e\sum_{\bfk\sigma\in\chi}\frac{|\Delta_\chi||t_\chi^{(0)}|^2}{2E_\bfk}
	[\calL_{\bfk-}(\omega+\Delta_{\bar\sigma0})
\nonumber\\&
	-\calL_{\bfk+}(\omega+\Delta_{2\sigma})],
\\
\Gamma_\chi(\mu_\chi)=&2e\sum_{k\sigma\in\chi}\frac{|\Delta_\chi||t_\chi^{(0)}|^2}{2E_k}
	[\calL_{\bfk-}^2(\omega+\Delta_{\bar\sigma0})
\nonumber\\&
	-\calL_{\bfk+}^2(\omega+\Delta_{2\sigma})],
\end{align}
\end{subequations}
respectively, where $\calL_{\bfk\pm}(\omega)=1/(\omega-[\mu_\chi\pm E_\bfk])$, $\bfk\in\chi$.

The quadratic dependence of the displacement $q$ in the current, Eq. (\ref{eq-ISmain}), is justified since the $q^2\ll q$ for small $q$. The quadratic component, thus, merely provides a minor modification to the linear displacement.

\subsection{Two-electron tunneling process}
\label{ssec-N02}
The dynamics of the average $N_{02}\equiv\av{\X{02}{}}$ is calculated through the equation of motion
\begin{align}
(i\dt-\Delta_{20})N_{02}=
	-\sum_{k\sigma}\sigma t_\chi\av{(\X{0\sigma}{}+\sigma\X{\bar\sigma2}{})\cs{k\bar\sigma}}e^{-i\mu_\chi t},
\end{align}
where the correlation function is treated by using perturbation theory, which to first order in $t_\chi$ gives
\begin{align}
\av{[(\X{0\sigma}{}+\sigma\X{\bar\sigma2}{})\cs{k\bar\sigma}](t)}=
\nonumber\\&\hspace{-3.5cm}=
	i\int_{-\infty}^tt_\chi(
		[G_{0\sigma}^>(t,t')
		+G_{\bar\sigma2}^>(t,t')]{\cal F}^<_{k\sigma\bar\sigma}(t',t)
\nonumber\\&\hspace{-3cm}
		-[G_{0\sigma}^<(t,t')+G_{\bar\sigma2}^<(t,t')]{\cal F}^>_{k\sigma\bar\sigma}(t',t)
		)e^{-i\mu_\chi t'}dt'.
\end{align}
The lesser and greater GFs for the QD, are here given by
\begin{subequations}
\begin{align}
G_{0\sigma}^<(t,t')=iN_\sigma e^{-i\Delta_{\sigma0}\tau},\ 
&
G_{0\sigma}^>(t,t')=-iN_0 e^{-i\Delta_{\sigma0}\tau},
\\
G_{\bar\sigma2}^<(t,t')=iN_2 e^{-i\Delta_{2\bar\sigma}\tau},\ 
&
G_{\bar\sigma2}^>(t,t')=-iN_{\bar\sigma} e^{-i\Delta_{2\bar\sigma}\tau},
\end{align}
\end{subequations}
where $N_p=\av{\X{pp}{}}$, denotes the occupation number for the QD state $\ket{p}$.
whereas the anomalous GFs for the leads are expressed as
\begin{align*}
{\cal F}_{k\sigma\bar\sigma}^{>}(t,t')\equiv&
	-i\av{\cc{k}(t)\cs{k\bar\sigma}(t')}
\nonumber\\=&
	i\sigma u_\bfk v_\bfk^*[f(E_\bfk)e^{iE_\bfk\tau}-f(-E_\bfk)e^{-iE_\bfk\tau}],
\\
{\cal F}_{k\sigma\bar\sigma}^{<}(t,t')\equiv&
	i\av{\cs{k\bar\sigma}(t')\cc{k}(t)}
\nonumber\\=&
	i\sigma u_kv_k^*[f(E_\bfk)e^{-iE_k\tau}-f(-E_\bfk)e^{iE_\bfk\tau}].
\end{align*}
The equation for $N_{02}$ is, thus, given by
\begin{align}
(i\dt-&\Delta_{20})N_{02}=-\sum_\chi\int\{(1+\alpha_\chi q)[(1+\alpha_\chi q)U_\chi(\omega,t')
\nonumber\\&
	-i\alpha_\chi\dot{q}V_\chi(\omega,t')]\}e^{-i\omega(t-t')}\frac{d\omega}{2\pi}dt'e^{-i(\omega_\chi t+\phi_\chi)},
\label{eq-N02}
\end{align}
where the amplitudes $U_\chi$ and $V_\chi$ are
\begin{subequations}
\begin{align}
U_\chi(\omega,t)=&
	-\sum_{\bfk\sigma\in\chi}\frac{|\Delta_\chi||t_\chi^{(0)}|^2}{2E_\bfk}
	[
	N_0\calL_{\bfk-}(\Delta_{\sigma0}-\omega)
\nonumber\\&
	+N_{\bar\sigma}\calL_{\bfk-}(\Delta_{2\bar\sigma}-\omega)
	+N_\sigma\calL_{\bfk+}(\Delta_{\sigma0}-\omega)
\nonumber\\&
	+N_2\calL_{\bfk+}(\Delta_{2\bar\sigma}-\omega)
	]
\\
V_\chi(\omega,t)=&\sum_{k\sigma\in\chi}\frac{|\Delta_\chi||t_\chi^{(0)}|^2}{2E_k}
	[
	N_0\calL_{\bfk-}^2(\Delta_{\sigma0}-\omega)
\nonumber\\&
	+N_{\bar\sigma}\calL_{\bfk-}^2(\Delta_{2\bar\sigma}-\omega)
	+N_\sigma\calL_{\bfk+}^2(\Delta_{\sigma0}-\omega)
\nonumber\\&
	+N_2\calL_{\bfk+}^2(\Delta_{2\bar\sigma}-\omega)
	].
\end{align}
\end{subequations}
Those amplitudes describe the cotunneling processes in which two electrons are either added or removed from the QD in absence and presence of the coupling to the vibrational mode, respectively.

\subsection{Quantum dot occupation numbers}
\label{ssec-QDoccupation}
It is clear that the supercurrent is to a great extent determined by the time-evolution of the electron occupation $N_p$ in the QD. We obtain those from the density matrix $\rho(t)=\{\av{\X{pp'}{}}(t)\}_{pp'}$ requiring $\sum_pN_p=1$. Within the employed level of approximation, $\rho$ is determined from the master equation $\dt{\bf N}=\sum_\chi t_\chi^2{\bf U}_\chi{\bf N}$, where ${\bf N}=(N_0\ N_\up\ N_\down\ N_2)^T$, whereas
\begin{align}
{\bf U}_\chi=
	\left(\begin{array}{cccc}
	\sum_\sigma\Lambda_{0\sigma}^\chi & \Gamma_{0\up}^\chi & \Gamma_{0\down}^\chi & 0 \\
	-\Lambda_{0\up}^\chi & -\Gamma_{0\up}^\chi+\Lambda_{\up2}^\chi & 0 & \Gamma_{\up2}^\chi \\
	-\Lambda_{0\down}^\chi & 0 & -\Gamma_{0\down}^\chi+\Lambda_{\down2}^\chi & \Gamma_{\down2}^\chi \\
	0 & -\Lambda_{\up2}^\chi & -\Lambda_{\down2}^\chi & -\sum_\sigma\Gamma_{\sigma2}^\chi
		\end{array}\right),
\end{align}
and where
\begin{subequations}
\begin{align}
\Gamma_{pq}^\chi=&
	2\pi\calN_\chi\Delta_{qp}^\chi
	\frac{\theta(\Delta_{qp}^\chi-|\Delta_\chi|)}{\sqrt{(\Delta_{qp}^\chi)^2-|\Delta_\chi|^2}},
\\
\Lambda_{pq}^\chi=&
	2\pi\calN_\chi\Delta_{qp}^\chi
	\frac{\theta(-\Delta_{qp}^\chi-|\Delta_\chi|)}{\sqrt{(\Delta_{qp}^\chi)^2-|\Delta_\chi|^2}}.
\end{align}
\end{subequations}
Here, $\calN_\chi$ is the density of electron states in lead $\chi$, whereas $\Delta_{pq}^\chi=\Delta_{pq}-\omega_\chi$.

\section{Dynamics of the quantum dot}
\label{sec-dynamics}
It can be seen in Tabs. \ref{tab-pd1} and \ref{tab-pd2}, and Fig. \ref{fig-pd} that there are several regimes in the $(\dote{0},U)$-space, in which the time-dependence of the quantum dot occupation numbers is very different. Here, we analyze each of the regimes in order to elucidate the characteristic time-scales involved in the electro-mechanical dynamics of the QD.

\begin{table}[t]
\caption{The three regimes in which at least one transition energy lies within the superconducting gap.}
\label{tab-pd1}
\begin{tabular}{l|c|c|c}
\hline\hline
 & I & II & III \\
 & $\Delta_{\sigma0}^\chi<-|\Delta_\chi|$
 & $|\Delta_{\sigma0}^\chi|<|\Delta_\chi|$
 & $|\Delta_{\sigma0}^\chi|<|\Delta_\chi|$
\\
 & $|\Delta_{2\sigma}^\chi|<|\Delta_\chi|$
 & $|\Delta_{2\sigma}^\chi|<|\Delta_\chi|$
 & $\Delta_{2\sigma}^\chi>|\Delta_\chi|$
\\
\hline
$\Gamma^\chi_{0\sigma}$ & 0 & 0 & 0 \\
$\Gamma^\chi_{\sigma2}$ & 0 & 0 & $>0$ \\
$\Lambda^\chi_{0\sigma}$ & $<0$ & 0 & 0 \\
$\Lambda^\chi_{\sigma2}$ & 0 & 0 & 0 \\
\hline\hline
\end{tabular}
\end{table}

\subsection{Regime I: Both transitions within the gap}
\label{ssec-RI}
Beginning with regime I, we find that the occupation numbers are constants of motion since the matrix $\bfU_\chi=0$, hence, $\dt\bfN=0$. The non-dynamical charge distribution on the QD, implies that the QD motion can be linearly excited by varying the bias voltage and superconducting phases. Integrating Eq. (\ref{eq-N02}), we obtain
\begin{align}
N_{02}=&
	\sum_\chi\Bigl\{
		\Bigl[
			(1+\alpha_\chi q)^2U_\chi'
			-\alpha_\chi
			\{
				(1+\alpha_\chi q)\dot{q}+\alpha_\chi\ddot{q}
			\}
			V_\chi''
\nonumber\\&\vphantom{\sum_\chi}
			-2\alpha_\chi^2\dot{q}^2U_\chi'''
		\Bigr]
				\cos(\omega_\chi t+\phi_\chi)
		-\alpha_\chi\dot{q}
		\Bigl[(1+\alpha_\chi q)V_\chi'
\nonumber\\&
		+2(1+\alpha_\chi q)U_\chi''
		-2\alpha_\chi\ddot{q}V_\chi'''
		\Bigr]
				\sin(\omega_\chi t+\phi_\chi)
		\Bigr\},
\end{align}
where $U_\chi'=U_\chi/(\Delta_{20}-\omega_\chi)$, $U_\chi''=U_\chi/(\Delta_{20}-\omega_\chi)^2$, and $U_\chi'''=U_\chi/(\Delta_{20}-\omega_\chi)^3$, and analogously for $V_\chi'$, $V_\chi''$, and $V_\chi'''$.

An order of magnitude estimate for the relative ratios between the parameters $J_S^\chi$, $\Gamma_S^\chi$, $U_\chi$, and $V_\chi$ gives
\begin{align}
\omega_\chi\frac{\Gamma_\chi}{J_\chi}\sim
	\left(\frac{eV}{|\Delta_\chi|}\right)^2,\ &
\omega_\chi\frac{V_\chi}{U_\chi}\sim
	1,\ &
\frac{e}{\hbar}\frac{U_\chi}{J_\chi}\sim
	\frac{eV}{|\Delta_\chi|}.
\end{align}
As we are considering small bias voltages, $eV\ll|\Delta_\chi|$, we proceed by neglecting the $\Gamma_\chi$ terms, and since we are considering weak coupling between the vibrational mode and the electronic degrees of freedom, we only keep terms at most linear in $\alpha_\chi$. The current, thus, reduces to
\begin{align}
\vphantom{\sum_\chi}
I_\chi(t)=&
	J_\chi
	\sum_{\chi'}\{(
		[1+2\alpha_\chi q+2\alpha_{\chi'}q]U_{\chi'}'
		-\alpha_{\chi'}\ddot{q}V_{\chi'}''
		)
\nonumber\\&\times\vphantom{\sum_{\chi'}}
		\cos(\omega_{\chi'}t+\phi_{\chi'})
	+\alpha_{\chi'}\dot{q}[2U_{\chi'}''-V_{\chi'}']
\nonumber\\&\times
	\sin(\omega_{\chi'}t+\phi_{\chi'})]
	\}
	\sin(\omega_\chi t+\phi_\chi),
\label{eq-JSchi}
\end{align}

\begin{table}[t]
\caption{The three regimes in which both transition energies lie outside the superconducting gap.}
\label{tab-pd2}
\begin{tabular}{l|c|c|c}
\hline\hline
 & IV & V & VI \\
 & $\Delta_{pq}^\chi<-|\Delta_\chi|$
 & $|\Delta_{\sigma0}^\chi|<-|\Delta_\chi|$
 & $\Delta_{pq}^\chi>|\Delta_\chi|$
\\
 & 
 & $|\Delta_{2\sigma}^\chi|>|\Delta_\chi|$
 & 
\\
\hline
$\Gamma^\chi_{0\sigma}$ & 0 & $<0$ & $>0$ \\
$\Gamma^\chi_{\sigma2}$ & 0 & $<0$ & $>0$ \\
$\Lambda^\chi_{0\sigma}$ & $<0$ & $>0$ & 0 \\
$\Lambda^\chi_{\sigma2}$ & $<0$ & $>0$ & 0 \\
\hline\hline
\end{tabular}
\end{table}

We note several things in this formula for the Josephson current. The first term would arise when the QD is rigid in space since there is no coupling to the vibrational motion. All other terms, which are proportional to $\alpha_\chi$, arise due to the movability of the QD in the Josephson junction. There are two terms directly depending on the position $q$ of the QD, as expected from previous studies.\cite{zhu2006,fransson2008} More interesting, however, is the dependence on the velocity $\dot{q}$ and acceleration $\ddot{q}$ of the QD, displayed in the fourth, fifth, and sixth terms. These dependences, hence, open possibilities for novel measurements and applications of shuttling QDs.

Following the procedure introduced in Refs. \onlinecite{zhu2006,fransson2008}, we derive the effective Hamiltonian $\Hamil_J$  by requiring that $2e\partial\Hamil_J/\partial\phi_\chi=I_\chi$, where $I_\chi$ is given in Eq. (\ref{eq-JSchi}). We briefly discuss the salient features of the result, while a detailed derivation will be presented elsewhere.
Within the given level of approximation, the Hamiltonian is linear in the coordinates $q,\dot{q},\ddot{q}$. Hence, the classical equation of motion from the Hamiltonian $\Hamil_{osc}=p^2/(2m)+k_cx^2/2+\Hamil_J$ is expected to be simple, in the sense that all features derived from the superconducting current act as a driving force $F(t)$. The equation of the QD motion will, therefore, be that of a driven oscillator, i.e. $m_c\ddot{q}+\gamma_N\dot{q}+k_cq=F(t)$, where the damping factor $\gamma_N$ contains the external damping due to mechanical friction. We notice \emph{en passant} that a quadratic approximation in $\alpha_\chi$ provides contributions that add a Josephson stiffness and damping. The linear spatial motion of the QD is, however, given by
\begin{align}
q(t)=&q_0\sin(\tilde{\omega}_0t+\delta_0)e^{-\gamma_Nt/2m}
	-q_0\sum_\chi\frac{1}{K_\chi}\biggl\{
		\frac{3\tilde{\alpha}_\chi}{4}U_\chi'
\nonumber\\&
		-\frac{\tilde{\alpha}_\chi}{4}
			[3U_\chi'-2\omega_\chi(2U_\chi''-V_\chi')]H(\omega_\chi,\phi_\chi)
\nonumber\\&
		+\sum_{\chi'\neq\chi}\biggl(
		\calQ_{\chi\chi'}(\omega_{\chi'})
		H(\omega_{\chi'},\phi_{\chi'})
		-\frac{1}{2}\calQ_{\chi\chi'}(\omega_{\chi'}-\omega_\chi)
\nonumber\\&\times
		H(\omega_{\chi'}-\omega_\chi,\phi_{\chi'}-\phi_\chi)
		-\frac{1}{2}\calQ_{\chi\chi'}(\omega_{\chi'}+\omega_\chi)
\nonumber\\&\times
		H(\omega_{\chi'}+\omega_\chi,\phi_{\chi'}+\phi_\chi)
		\biggr)\biggr\},
\label{eq-q}
\end{align}
where
\begin{align*}
\calQ_{\chi\chi'}(\omega,\phi)=&
	[\tilde{\alpha}_\chi+2\tilde{\alpha}_{\chi'}]U_{\chi'}'
		+\tilde{\alpha}_{\chi'}\omega[2U_{\chi'}''-V_{\chi'}'],
\\
H(\omega,\phi)=&\frac{1-(\omega/\omega_0)^2}{[1-(\omega/\omega_0)^2]^2+[\gamma_N\omega/k_c]^2}
\nonumber\\&\times
	\biggl[
	\cos(\omega t+\phi)
	+\frac{\gamma_N\omega}{k_c}
		\frac{\sin(\omega t+\phi)}{1-(\omega/\omega_0)^2}
	\biggr],
\end{align*}
whereas $\tilde{\alpha}_\chi=q_0\alpha_\chi$ and $K_\chi=k_cq_0^2/E_J^\chi$ are dimensionless parameters, whereas $E_J^\chi=J_\chi/2e$ is the Josephson energy. The first term in Eq. (\ref{eq-q}) describes the unperturbed damped motion of the QD around its equilibrium position $q_\text{eq}$, where $q_0$ and $\delta_0$ are to be determined from the initial conditions, whereas $\tilde{\omega}_0=\sqrt{\omega_0^2-[\gamma_N/2m]^2}$ ($\omega_0=\sqrt{k_c/m_c}$) is the eigenfrequency of the damped (undamped) mechanical oscillations. The only time-scales introduced in the Josephson current are the ones associated with the eigenfrequency $\tilde{\omega}_0$ and the Josephson frequencies $\omega_\chi$, which is contrasted by the properties of the other regimes, as we shall see below.

This extended analysis of the expected Josephson current in term of the QD motion is possible only since the occupation numbers are constants of motion. Hence, for the remaining regimes we will be content with establishing the characteristic time-scales.

\subsection{Regimes II and III: Single transition within the gap}
\label{ssec-RII}
Within regime II, the parameters $\Gamma_{pq}^\chi=0$, for all transitions $pq$, and also $\Lambda_{\sigma2}^\chi=0$ since $\Delta_{2\sigma}^\chi>-|\Delta_\chi|$, whereas $\Lambda_{0\sigma}^\chi\neq0$ since $\Delta_{\sigma0}^\chi<-|\Delta_\chi|$. Hence, the number $N_2$ is a constant of motion, while the remaining system of equations has the solution
\begin{subequations}
\begin{align}
N_0(t)=&
	\tilde{N}_0\prod_{\chi\sigma}e^{\Lambda_{0\sigma}^\chi\int_{t_0}^tt_\chi^2(t')dt'},
\label{eq-N0I}
\\
N_\sigma(t)=&
	-\sum_\chi\Lambda_{0\sigma}^\chi\int_{t_0}^tt_\chi^2(t')N_0(t')dt',
\label{eq-NsI}
\end{align}
\end{subequations}
for some initial time $t_0$ at which $\tilde{N}_0=N_0(t_0)$. Under the local approximation of the tunneling rate, $N_0$ acquires the time-evolution
\begin{align*}
	\prod_{\chi\sigma}
	e^{
		\Lambda_{0\sigma}^\chi(t_\chi^{(0)})^3\alpha_\chi\dot{q}
		\{
		[1+\alpha_\chi q]
		[1-\alpha_\chi\dot{q}(t-t_0)]
		+\alpha_\chi^2\dot{q}^2(t-t_0)^2/3
		\}
		(t-t_0)
		}
		.
\end{align*}
The exponent periodically changes sign with the QD velocity $\dot{q}$, and position $q$, which provides an oscillatory behavior of the occupation numbers $N_0$ as well as for $N_\sigma$. Physically, this means that the electron density in the QD periodically grows and wanes as the QD moves, which opens for possibilities to load and unload electron density on the QD at different leads, or, single electron shuttling between the superconducting leads via the QD. The properties of the QD in regime III are analogously obtained by interchanging the roles of $N_0$ and $N_2$.

The time-scale $\tau$ for loading (unloading) electron density on the QD is related to the transition energy $\Delta_{\sigma0}^\chi$ ($\Delta_{2\sigma}^\chi$), and it can be noted that $\tau\rightarrow0$ as $\Delta_{\sigma0}^\chi\rightarrow-|\Delta_\chi|$ ($\Delta_{2\sigma}^\chi\rightarrow|\Delta_\chi|$), while it is determined solely by the density of electron states, $\tau\sim1/N_\chi$, in the leads for transition energies far below (above) the superconducting gap. The dynamics of the occupation numbers introduce an additional time-scale to the ac Josephson current which is associated with the energies of the QD transitions, and which is different from the one introduced through the Josephson frequency.

\subsection{Regime IV and VI: Both transitions below or above the gap}
\label{ssec-RIV}
In regime IV, all transition energies $\Delta_{pq}<-|\Delta_\chi|$, which implies that $\Gamma_{pq}^\chi=0$, whereas $\Lambda_{pq}^\chi\neq0$. We can, then, integrate the occupation of the empty state according to Eq. (\ref{eq-N0I}), whereas the occupations for the one- and two-electron states are given by
\begin{subequations}
\begin{align}
N_\sigma(t)=&
	-\prod_\chi\int_{t_0}^t
		e^{\Lambda_{\sigma2}^\chi\int_{t'}^t t_\chi^2(t'')dt''}
		\sum_{\chi'} t_{\chi'}^2(t')\Lambda_{0\sigma}^{\chi'}N_0(t')
	dt',
\label{eq-N1IV}
\\
N_2(t)=&
	-\sum_\sigma\Lambda_{\sigma2}\int_{t_0}^tN_\sigma(t')dt'.
\end{align}
\end{subequations}
While both the empty and one-electron states, again, depend on the periodic motion and velocity of the QD, the two-electron state also acquires an oscillatory behavior since it depends on the integrated time-evolution of the occupation of the other states. It is noticeable, however, that the time-evolution of the electron occupation in the QD here is related to the (four) time-scales associated with the transition energies $\Delta_{\sigma0}$ and $\Delta_{2\sigma}$. In particular, the one-electron occupations $N_\sigma$ strongly depend on the rate of the transitions $\X{0\sigma}{}$ and $\X{\sigma2}{}$. The properties of regime VI are obtained by noticing that all $\Gamma_{pq}\neq0$ and all $\Lambda_{pq}=0$ such that the roles of $N_0$ and $N_2$ become interchanged.

\subsection{Regime V: Gap between the lower and upper transitions}
\label{ssec-RV}
In regime V, finally, $\Delta_{\sigma0}^\chi<-|\Delta_\chi|$ and $\Delta_{2\sigma}^\chi>|\Delta_\chi|$, which leads to that $\Lambda_{0\sigma}^\chi<0$, $\Gamma_{\sigma0}^\chi=0$, $\Lambda_{\sigma2}^\chi=0$, and $\Gamma_{\sigma2}^\chi>0$. $N_0$ and $N_2$ are, thus, determined by Eq. (\ref{eq-N0I}), requiring the replacements $\Lambda_{0\sigma}^\chi\rightarrow\Gamma_{\sigma2}^\chi$ and $\tilde{N}_0\rightarrow\tilde{N}_2$ in the expression for $N_2$. It is understood that $N_0$ decreases (increases) while $N_2$ increases (decreases), which is expected from conservation of probability. The one-electron occupations $N_\sigma$ depend on the integrated time-evolution of $N_0$ and $N_2$, that is,
\begin{align}
N_\sigma(t)=&
	-\sum_\chi
	\int_{t_0}^t
		t_\chi^2(t')[\Lambda_{0\sigma}^\chi N_0(t')-\Gamma_{\sigma2}^\chi N_2(t')]
	dt'
	.
\end{align}
The occupation of the one-electron states can, thus, be viewed as resulting from the imbalance between the occupation in the empty and two-electron states

\section{Summary and conclusions}
\label{sec-discussion}
We have studied the dynamics of a single level molecular QD embedded in a Josephson junction, in which the mechanical motion of the QD is couples to the supercurrent. In the static limit, i.e. when the QD is rigid, the supercurrent between the superconducting electrodes is mediated via two-electron transitions in the QD, a behavior which remains also when the QD moves between the electrodes. The rate of the two-electron transitions naturally depends on the occupation of the QD states and the energies of the transitions between the states. The rate does, in addition, explicitly depend on the motion of the QD.

Our main focus has, in the present paper, been devoted to extract the time-scales that are involved in the dynamics of the QD occupation numbers. It turns out that we can clearly distinguish between six separate regimes in the phase space of the single electron energy level $\dote{0}$ and the QD charging energy $U$, see the phase diagram in Fig. \ref{fig-pd}. The boundaries between the various regimes are set by pairing potential (gap function) of the electrodes, and the dynamics of the QD occupation depends on whether the single-electron transitions in the QD lie within or outside the gap.

In all regimes but one, there are more than two time-scales involved in the dynamics of the QD occupation. Those additional time-scales are typically defined by the energies of the single-electron transitions relative to the superconducting gap (and the chemical potential). For transition energies very far below (above) the gap, the time-scale for occupying, or deoccupying, the corresponding states are set by the density of electron states in the electrodes, while the time-scale tends to zero as the transition energies approach the edge of the gap from below (above). As the transition energies lie within the gap function, the occupation numbers of the QD are constants of the motion which implies that the dynamics of the QD occupation depends solely on i) the two basic time-scales set by the bias voltage (Josephson frequency) and the eigenfrequency $\tilde{\omega}_0$ of the oscillator, and ii) the phase difference between the electrodes.

An experimental set-up to test our predictions would be possible by having gap functions $|\Delta|\sim10$ meV, which is pertinent to MgB$_2$,\cite{stipe2001} in order to obtain a sufficiently small mechanical damping. By also acquiring a vibrational frequency of $\omega_0/2\pi\sim1$ GHz,\cite{huang2003} one should be able to tune the Josephson frequency such that $\omega_0/\omega_J\gtrsim0.1$. A coupling strength of \cite{irish2005} $\alpha/\omega_0\sim10^{-1}\ -\ 10^{-3}$ should be sufficient for an efficient read-out. Finally, the charging energy $U$ of the QD may be of the order of 1 | 10 meV, while the level spacing should preferably be larger than $U$ in order to control the electron occupation. For this requirement on the charging energy, one would have access to the regimes I | IV and VI, whereas the regime V would become accessible by requiring a charging energy $U>2|\Delta|$.

J.F. acknowledges support from the Swedish Research Council and Royal Swedish Academy of Sciences. A.V.B. and J.X.Z. acknowledge that this work was supported by US DOE, LDRD and BES, and was carried out under the auspices of the NNSA of the US DOE at LANL under Contract No. DE-AC52-06NA25396.

\end{document}